\documentclass[12pt, draftclsnofoot, onecolumn,romanappendices]{IEEEtran}
\usepackage{bm,cite}
\usepackage{amsmath}
\usepackage{amsthm}
\usepackage{amsbsy}
\usepackage{epsfig}
\usepackage{latexsym}
\usepackage{amssymb} 
\usepackage{wasysym}
\usepackage{placeins}  
\usepackage{color}
\usepackage{rotating}   
\usepackage[usenames,dvipsnames]{xcolor}
\usepackage{dsfont}
\usepackage{caption}
\usepackage{subcaption}
\usepackage{algorithm}
\usepackage{algorithmic}
\usepackage{setspace}

\usepackage{tikz}
  
\usetikzlibrary{shapes,arrows,snakes}
\usetikzlibrary{calc}

\IEEEoverridecommandlockouts
\DeclareMathAlphabet{\mathbit}{OML}{cmr}{bx}{it}
\DeclareMathAlphabet{\mathsf}{OT1}{cmss}{m}{n}
\DeclareMathAlphabet{\mathTXf}{OT1}{cmss}{bx}{it}

\DeclareMathOperator{\diag}{diag}

\DeclareMathOperator{\trace}{tr}

\DeclareMathOperator*{\argmin}{argmin}
\DeclareMathOperator*{\argmax}{argmax}

\DeclareMathOperator{\Rate}{R}

\DeclareMathOperator{\Opt}{\mathrm{opt}}

\DeclareMathOperator{\Tot}{tot}

\newcommand{\bE}{\mathbf{E}} 
 
\newcommand{\bG}{\mathbf{G}} 
\newcommand{\bH}{\mathbf{H}} 
\newcommand{\bI}{\mathbf{I}}

\newcommand{\bM}{\mathbf{M}}

\newcommand{\bT}{\mathbf{T}} 
 
\newcommand{\bV}{\mathbf{V}} 
\newcommand{\bW}{\mathbf{W}} 
\newcommand{\bX}{\mathbf{X}}

\newcommand{\bd}{\bm{d}} 
\newcommand{\be}{\bm{e}}

\newcommand{\bs}{\bm{s}}

\newcommand{\by}{\bm{y}}

\newcommand{\LB}{\left(}
\newcommand{\RB}{\right)}

\newcommand{\I}{\mathbf{I}} 

\newcommand{\Fro}{{\mathrm{F}}}

\newcommand{\BR}{{\text{BR}}}

\newcommand{\E}{{\mathrm{E}}}

\newcommand{\trans}{{\text{T}}} 
 
\newcommand{\He}{{{\mathrm{H}}}}

\newcommand{\In}{{\text{in}}}

\newcommand{\xv}{\mathbf{x}}

\theoremstyle{remark}
\newtheorem{remark}{Remark} 
\theoremstyle{example}

\begin{document} 
\title{Robust Precoding for Network MIMO with Hierarchical CSIT} 
\author{\IEEEauthorblockN{Paul de Kerret\IEEEauthorrefmark{1}, Richard Fritzsche\IEEEauthorrefmark{4}, David Gesbert\IEEEauthorrefmark{1}, and Umer Salim\IEEEauthorrefmark{3}} 

\IEEEauthorblockA{\IEEEauthorrefmark{1}
Mobile Communication Department, Eurecom, {\tt{\{dekerret,gesbert\}@eurecom.fr}}}

\IEEEauthorblockA{\IEEEauthorrefmark{4}TU Dresden, Vodafone Chair Mobile Communications Systems, {\tt{richard.fritzsche@tu-dresden.de}}}

\IEEEauthorblockA{\IEEEauthorrefmark{3}Intel Mobile Communications, {\tt{umer.salim@intel.com}}}
}


\maketitle

\begin{abstract} 
In this work\footnote{The research leading to these results has received funding from the European Union Seventh Framework Programme (FP7/2007-2013) under grant agreement n° 317941. The authors would like to acknowledge the contributions of their colleagues in iJOIN, although the views expressed are those of the authors and do not necessarily represent the project.} we consider a wireless network with~$K$ cooperating transmitters (TXs) serving jointly $K$~receivers (RXs). Due to the practical limitations of the backhaul network, it is relevant to consider a setting where each TX receives its own imperfect estimate of the multi-user channel state, denoted as the distributed channel state information (CSI) setting. We focus in this work on a particular distributed CSI configuration called \emph{hierarchical CSI} configuration in which the TXs can be ordered by increasing level of CSI. This scenario is particularly relevant for future networks with heterogeneous backhaul where the TXs connected with a weak backhaul link will receive only a coarse estimate while the TXs with a stronger backhaul will have a more accurate CSI. In that scenario, we formulate the optimal precoding as a \emph{team decision} problem. Solving optimally this problem is extremely challenging such that we propose a heuristic approach allowing to obtain a simple, yet efficient and practical, precoding algorithm. The proposed precoding algorithm exploits the hierarchical structure of the CSI to make the transmission more robust to the imperfect CSI knowledge at the TXs.
\end{abstract}
\IEEEpeerreviewmaketitle

\section{Introduction}

Network (or Multi-cell) MIMO methods, whereby multiple interfering TXs share user messages and allow for joint precoding, are currently considered for next generation wireless networks \cite{Gesbert2010}. With perfect message and CSI sharing, the different TXs can be seen as a unique virtual multiple-antenna array serving all RXs in a multiple-antenna broadcast channel (BC) fashion and well known precoding algorithms can be used \cite{Christensen2008}. However, this requires the feedback of a very accurate multi-user CSI to the TX side to achieve the desired high performances\cite{Caire2010}. As a consequence, there has been a large amount of works dealing with the feedback of the CSI (See \cite{Love2008} and references therein) and the design of robust precoders (See for example \cite{Vucic2009,Shenouda2007,Negro2012,Fritzsche2013a}).

However, the large literature on robust precoding typically assumes \emph{centralized} CSIT, i.e., that the precoding is based on the basis of a \emph{single} imperfect channel estimate. It implies that either the precoding is done in a central node or the CSI is \emph{perfectly shared} between the TXs. Although meaningful in the single TX case with multiple-antennas, this assumption is often unrealistic for distant cooperating TXs where the CSI obtained locally has to be exchanged between the distant TXs. This step introduces inevitably some delay and may also require further quantization. Hence, it is practically relevant for joint precoding across distant TXs to allow for the case where each TX receives its \emph{own} channel estimate, which we denote as the \emph{distributed CSI} configuration \cite{dekerret2013_thesis}.

With distributed CSI, the design of a joint precoder is particularly challenging and only few results are available. This problem belongs in fact to the category of \emph{Team Decision} problems \cite{Radner1962,Marschak1972,Ho1980} for which only particular cases have been solved. In \cite{dekerret2012_TIT}, the number of Degrees-of-freedom (DoF) obtained with conventional ZF precoding is derived and some precoding schemes improving the DoF are given. In \cite{Zakhour2010a}, a robust precoding algorithm is designed for the case of two TXs having distributed CSIT. However, the algorithm developed is computationally demanding and does not provide any insight. In \cite{Fritzsche2013b}, it is shown how to refine the precoder when more accurate CSI is available locally. However, this approach can only be used for a specific CSI configuration.

In this work, we consider a CSI configuration, denoted as \emph{hierarchical} (also called ``nested" \cite{Ho1972}) where the TXs can be ordered in such a way that a TX~$j$ has a more accurate channel estimate than a TX~$k$, if $k<j$. Specifically, our main contributions read as follows:
\begin{itemize}
\item We formulate the problem of precoding with hierarchical CSI. Although the hierarchical structure allows to simplify the optimization, it remains a difficult stochastic optimization problem \cite{Kall1994}. 
\item Thus, we propose a simple robust hierarchical precoding algorithm exploiting the hierarchical structure of the CSI. The proposed algorithm outperforms precoding schemes from the literature. Furthermore, the approach developed can potentially be adapted to many other scenarios.
\end{itemize}

\section{System Model} 

\subsection{Received Signal}
We study the transmission from $K$~TXs to $K$~RXs where the $i$-th TX is equipped with $M_i$~antennas and transmits $d_i$ streams to the $i$-th RX equipped with $N_i$ antennas. The total number of RX antennas, the total number of TX antennas and the total number of streams are respectively given by
\begin{equation}
N_{\Tot}\triangleq \sum_{i=1}^K N_i,\quad  M_{\Tot}\triangleq \sum_{i=1}^K M_i,\quad d_{\Tot}\triangleq \sum_{i=1}^K d_i.
\label{eq:SM_1}
\end{equation}  
We further assume that the RXs have perfect CSI as we decide to focus primarily on the challenge of conveying CSI back to the TXs through some form of feedback. We consider that linear filtering are used and that the RXs treat interference as noise. The channel from the $K$~TXs to the $K$~RXs is represented by the channel matrix~$\mathbf{H}^{\He} \in \mathbb{C}^{N_{\Tot}\times M_{\Tot}}$ where $\bH_{ik}^{\He}\in \mathbb{C}^{N_i\times M_k}$ denotes the channel matrix from TX~$k$ to RX~$i$ and has all its elements i.i.d. as $\mathcal{CN}(0,\rho_{i,k}^2)$ and independent of the other channel matrices. 

The transmission is then described as
\begin{equation}
\begin{bmatrix}
\by_1\\\vdots\\\by_K
\end{bmatrix}
=
\mathbf{H}^{\He}\xv
+
\bm{\eta}
=
\begin{bmatrix}
\bH^{\He}_1\xv\\\vdots\\
\bH^{\He}_K\xv
\end{bmatrix}
+
\begin{bmatrix}
\bm{\eta}_1\\\vdots\\
\bm{\eta}_K
\end{bmatrix}
\label{eq:SM_2}
\end{equation}
where $\by_i\in \mathbb{C}^{N_i\times 1}$ is the signal received at the $i$-th RX, $\bH^{\He}_i \in \mathbb{C}^{N_i\times M_{\Tot}}$ the channel from all TXs to the $i$-th RX, and $\bm{\eta}\triangleq[\bm{\eta}_1,\ldots,\bm{\eta}_K]^{\trans}\in \mathbb{C}^{N_{\Tot}\times 1}$ the normalized Gaussian noise with its elements i.i.d. as $\mathcal{CN}(0,1)$. 

The multi-TX transmitted signal~$\xv\!\in \!\mathbb{C}^{M_{\Tot}\times 1}$ is obtained from the symbol vector $\bs\!\triangleq\![\bs_1^{\trans},\ldots,\bs_K^{\trans}]^{\trans}  \!\in\!  \mathbb{C}^{d_{\Tot}\times 1}$ with its elements i.i.d. $\mathcal{CN}(0,1)$ as
\begin{equation}
\xv=\bT \bs=
\begin{bmatrix}
\bT_1,
\hdots,
\bT_K
\end{bmatrix}
\begin{bmatrix}
\bs_1\\
\vdots\\
\bs_K
\end{bmatrix} =\sum_{k=1}^K\bT_k\bs_k
\label{eq:SM_3}
\end{equation} 
with $\bT_j\in \mathbb{C}^{M_{\Tot}\times d_j}$ being the precoder serving user~$j$ and $\bT \in \mathbb{C}^{M_{\Tot}\times d_{\Tot}}$ being the multi-user precoder. We also introduce the matrix~$\bW_j^{\He}\in \mathbb{C}^{M_j\times d_{\Tot}}$ to denote the precoding coefficients at TX~$j$ such that the signal transmitted by TX~$j$, denoted by~$\xv_{j}\!\in \!\mathbb{C}^{M_{j}\times 1}$, is given by
\begin{equation}
\xv_j=\bW_j^{\He}\bs.
\label{eq:SM_4}
\end{equation} 
The multi-user precoder~$\bT$ is then alternatively written as
\begin{equation}
\bT=
\begin{bmatrix}
\bW_1^{\He}\\
\vdots\\
\bW_K^{\He}
\end{bmatrix}.
\label{eq:SM_5}
\end{equation} 

Finally, the received signals are further processed by the RX filter~$\bG^{\He} \in \mathbb{C}^{d_{\Tot}\times N_{\Tot}}$ equal to
\begin{equation}
\bG^{\He}\triangleq
\mathrm{blockdiag}(\bG^{\He}_1,\ldots,\bG_K^{\He}) 
\label{eq:SM_6}
\end{equation} 
with $\bG_k^{\He}\in \mathbb{C}^{d_k\times N_k}$ being the RX filter at RX~$k$. It follows that we can define the mean square error (MSE) matrix at RX~$k$ for given precoders and RX filters, denoted by~$\bM_k\!\in\! \mathbb{C}^{d_k\times d_k}$, as
\begin{equation}
\begin{aligned}
\bM_k&\triangleq \E_{\bd_k}[(\bd_k-\bG_k^{\He}\by_k)(\bd_k-\bG_k^{\He}\by_k)^{\He}]\\
&= \I_{N_k}+\bG_k^{\He}\bG_k+\bG_k^{\He}\bH^{\He}_k\bT\bT^{\He}\bH_k\bG_k\\
&~~~~-\bG_k^{\He}\bH_k^{\He}\bT_{k}-\bT_k^{\He}\bH_k\bG_{k}.
\end{aligned}
\label{eq:SM_7}
\end{equation}    
Following the assumption of Gaussian signaling, the rate of user~$i$ can be written as \cite{Cover2006}
\begin{equation}
R_k\triangleq \log_2\left| \bM_k^{-1}\right|,\qquad \qquad \forall k\in \{1,\ldots,K\}.
\label{eq:SM_8}
\end{equation}
Finally, we define the sum rate~$R$, which will be our main figure-of-merit, as
\begin{equation}
R\triangleq \sum_{k=1}^KR_k.
\label{eq:SM_8}
\end{equation}


\subsection{Distributed Precoding and Distributed CSIT}
In the distributed CSIT model considered here, each TX receives its own CSI based on which it designs its transmission parameters without any additional communication to the other TXs\cite{dekerret2012_TIT,dekerret2013_thesis}. The actual exchange mechanism based on which the TXs receive the multi-user channel estimate is out of the scope of its paper and a research topic in its own right.

TX~$j$ has then the knowledge of the global multi-user channel estimate~$(\hat{\bH}^{(j)})^{\He}\in \mathbb{C}^{N_{\Tot}\times M_{\Tot}}$. We define~$(\hat{\bH}^{(j)}_{i})^{\He}\in \mathbb{C}^{N_{i} \times M_{\Tot}}$ in a similar fashion as its counterpart~$\bH_{i}^{\He}\in \mathbb{C}^{N_{i} \times M_{\Tot}}$ with perfect CSIT.

Hence, TX~$j$ designs its transmit coefficient~$\xv_j\in \mathbb{C}^{M_j\times 1}$ as a function of~$\hat{\bH}^{(j)}$. The transmit signal~$\xv_j$ is then given by
\begin{equation}
\xv_j=\bW_j^{\He}\!(\hat{\bH}^{(j)})\bs
\label{eq:SM_9}
\end{equation}

Due to the assumption of distributed precoding, the actual precoder used for the transmission is equal to
\begin{equation}
\bT \triangleq 
\begin{bmatrix}
\bW_1^{\He}\!(\hat{\bH}^{(1)})\\
\bW_2^{\He}\!(\hat{\bH}^{(2)})\\
\vdots\\
\bW_K^{\He}\!(\hat{\bH}^{(K)})
\end{bmatrix}.
\label{eq:SM_10}
\end{equation}

\subsection{Hierarchical Channel State Information}\label{se:SM:DCSI}

TX~$j$ receives an estimate~$\hat{\bH}^{(j)}$ of the multi-user channel~$\bH$. The estimate $\hat{\bH}^{(j)}$ can take a priori any form depending on the transmission scenarios considered, and we focus in this work on a particular CSI configuration called the \emph{hierarchical CSI configuration}. 

In the hierarchical CSI configuration, the TXs can be ordered by increasing quality of CSI, i.e., such that the estimate of TX~$j$ is ``included" in the estimate of TX~$j\!+\!1$. This scenario is in particular obtained if a multi-level quantization scheme is used \cite{Ng2009,Zakhour2010a}. In this quantization scheme, the same codebook is used for all the TXs and each TX decodes the estimate up to a number of bits corresponding to the quality of its feedback channel. By decoding less bits, TX~$j$ is then able to reconstitute the channel estimate at TX~$k$, for $k<j$~\cite{Ho1972}.

In particular, we will model the imperfect estimate~$\hat{\bH}^{(j)}$ at TX~$j$ as
\begin{equation}
\{\hat{\bH}^{(j)}\}_{i,k}=\sqrt{1-(\sigma_{i,k}^{(j)})^2}\{\hat{\bH}\}_{i,k}+\sigma_{i,k}^{(j)}\{\bm{\Delta}\}^{(j)}_{i,k},\quad \forall i,k
\label{eq:SM_11}
\end{equation}
where $\{\bm{\Delta}\}^{(j)}_{i,k}\sim\mathcal{CN}(0,1)$ and $\sigma_{i,k}^{(j)}\in (0,1), \forall k,i$ represents the quality of the CSIT at TX~$j$. To model the hierarchical quantization, we hence assume that TX~$j$ has access to $\hat{\bH}^{(k)}$ for $k\leq j$.

\begin{remark}
Any other model for the imperfect knowledge of the channel state can be used in our approach. It is only critical to have the \emph{hierarchical} structure of the CSI. \qed
\end{remark}

\subsection{Precoding with Distributed CSIT: Team Decision Problem}\label{se:SM:TD}
With distributed CSIT, the precoding problem can then be formulated as the following Team Decision problem \cite{Radner1962,Marschak1972,Ho1980}:
\begin{equation}
\begin{aligned}
&(\bW_1^{\star},\ldots,\bW_K^{\star})\\
&=\argmax_{(\bW_1,\ldots,\bW_K)} \E[\Rate(\bW_1(\hat{\bH}^{(1)}),\ldots,\bW_K(\hat{\bH}^{(K)}))] \\
&~~~~~~~~~\text{, s.to $\|\bW_j(\hat{\bH}^{(j)})\|_{\Fro}^2\leq P_j,\forall j\in \{1,\ldots, K\}$.}  
\end{aligned}
\label{eq:optimization_opt}
\end{equation} 
A necessary condition for any optimal precoding strategy is that it should also be a \emph{best-response} strategy. This means that each TX applies the best strategy \emph{given the strategies of the others TXs}\cite{Nash1951}. Mathematically, a best-response power allocation policy~$(\bW_1^{\BR},\ldots,\bW_K^{\BR})$ satisfies, $\forall j$,
\begin{equation}
\begin{aligned}
&\bW_j^{\BR}(\hat{\bH}^{(j)})\\
&=\argmax_{\bW_j} \E_{|\hat{\bH}^{(j)}}[\Rate(\bW^{\BR}_1\!\!,\!\ldots,\bW^{\BR}_{j-1},\!\bW_j,\bW^{\BR}_{j+1},\!\ldots,\!\!\bW^{\BR}_K)] \\
&~~~~~~~~~~~~~~~~~~~~~~~~~~~~~~~  \text{, s.to $\|\bW_j(\hat{\bH}^{(j)})\|_{\Fro}^2\leq P_j$.}  
\end{aligned}
\label{eq:optimization_best}
\end{equation} 
Solving the best-response optimization is usually more intuitive and more tractable. It corresponds however only to a local optimum of the original optimization problem.

Coming back to hierarchical CSI, the fundamental property of the hierarchical CSI configuration lies in the fact that TX~$j$ is able to carry out the signal processing which was done at TX~$k$ for $k<j$ to obtain the precoding decision~$\bW_k(\hat{\bH}^{(k)})$. Hence, the precoding decisions~$\bW(\hat{\bH}^{(k)})$ for $k<j$ are already given when considering the optimization at TX~$j$. The best-response optimization problem is then simplified to
\begin{equation}
\begin{aligned}
&\bW_j^{\BR}(\hat{\bH}^{(j)})\\
&=\argmax_{\bW_j} \E_{|\hat{\bH}^{(j)}}[\Rate(\bW_j,\bW^{\BR}_{j+1}(\hat{\bH}^{(j+1)})\!,\!\ldots\!,\!\bW^{\BR}_K(\hat{\bH}^{(K)}))] \\
&~~~~~~~~~~~~~~~~~~~~ \text{, s.to $\|\bW_j(\hat{\bH}^{(j)})\|_{\Fro}^2\leq P_j$.}  
\end{aligned}
\label{eq:optimization_hierarchical}
\end{equation} 
Yet, this remains a difficult problem as it requires to \emph{estimate} the precoding decisions at the TX~$k$ for $k=j+1,\ldots,K$. Solving this problem optimally was not possible and we propose in the following a simple, yet efficient, heuristic precoding algorithm.

\section{Hierarchical Precoding Algorithm}


\subsection{Hierarchical Precoding Algorithm}
Even though TX~$j$ does not know the information obtained at TX~$k$ for $k>j$, it can use the statistical information (available at every TX) to obtain an estimate of the precoding strategy which will be used at TX~$k$ for $k>j$. Based on this statistical information, each TX should optimize the conditional expectation as in \eqref{eq:optimization_hierarchical}. This is a functional stochastic optimization problem \cite{Kall1994} and is out of the scope of this paper. 

As a consequence, we use in the following the simplifying assumption that TX~$j$, when computing its precoding coefficient, implicitly assumes that TX~$k$ with $k>j$ shares the same channel estimate than he does (while TX~$k$ for $k>j$ has in fact a more accurate one, as described in Subsection~\ref{se:SM:DCSI}). Following this approximation, the optimization problem \eqref{eq:optimization_hierarchical} simplifies to
\begin{equation}
\begin{aligned}
&(\bW_j,\bV_{j+1}\ldots,\bV_K)\\
&=\argmax_{(\bW_j,\ldots,\bW_K)} \E[\Rate(\bW_j(\hat{\bH}^{(j)}),\ldots,\bW_K(\hat{\bH}^{(j)}))] \\
&~~~~~~~~~\text{, s.to $\|\bW_k(\bH^{(j)})\|_{\Fro}^2\leq P_k, \forall k\in \{j,\ldots, K\}$.}  
\end{aligned}
\label{eq:optimization_approximation}
\end{equation}   
\begin{remark}
Only~$\bW_j^{\He}(\bH^{(j)})$ is effectively used for the transmission since TX~$k$ with $k>j$ will use the more accurate information available locally to improve the precoding decision. This is why we have introduced the notation~$\bV_k$ to denote the precoding coefficients which will not be effectively used in the transmission. \qed
\end{remark}

\subsection{Precoding Algorithm for~$\bW_j^{\He}(\bH^{(j)})$}
We consider without loss of generality the optimization at TX~$j$. The precoding coefficients~$\bW_k^{\He}$ for $k<j$ are not part of the optimization problem since there have been already determined by the TXs having a less accurate channel estimate. Indeed, following the hierarchical CSI configuration, TX~$j$ can reconstitute the precoding decisions~$\bW_{k}^{\He}$ taken at TX~$k$ for $k<j$. Let us first introduce $M_{\In}$ and $M_{\Opt}$ as respectively the number of antennas with given precoding coefficients and the number of antennas with undetermined precoding coefficients:
\begin{equation}
\begin{aligned} 
M_{\In}\triangleq \sum_{k=1}^{j-1}M_k,\qquad M_{\Opt}\triangleq \sum_{k=j}^{K} M_k.
\end{aligned} 
\end{equation}   
The multi-user precoder can then be written as
\begin{equation}
\bT=\begin{bmatrix}
\bW^{\He}_{\In}\\
\bW^{\He}_{\Opt}
\end{bmatrix}
\label{eq:Algo_2}
\end{equation} 
where~$\bW^{\He}_{\In}\in \mathbb{C}^{M_{\In}\times d_{\Tot}}$ and~$\bW^{\He}_{\Opt}\in \mathbb{C}^{M_{\Opt}\times d_{\Tot}}$ are defined as
\begin{equation}
\bW^{\He}_{\In}\triangleq \begin{bmatrix}
\bW_1^{\He}\\
\vdots\\
\bW_{j-1}^{\He}
\end{bmatrix},\qquad \bW^{\He}_{\Opt}\triangleq \begin{bmatrix}
\bW_j^{\He}\\
\vdots\\
\bW_K^{\He}
\end{bmatrix} . 
\label{eq:Algo_3}
\end{equation}
The precoding coefficients in~$\bW^{\He}_{\In}$ are fixed such that it remains only to maximize the sum rate according to~$\bW^{\He}_{\Opt}$. Following the same idea, we also split the multi-user channel~$\bH^{\He}$ into the two parts~$\bH_{\In}^{\He}\in \mathbb{C}^{N_{\Tot}\times M_{\In}}$ and $\bH_{\Opt}^{\He}\in \mathbb{C}^{N_{\Tot}\times M_{\Opt}}$ such that
\begin{equation} 
\bH^{\He}=\begin{bmatrix}\bH_{\In}^{\He}&\bH_{\Opt}^{\He}\end{bmatrix}.
\label{eq:Algo_4}
\end{equation}
In the following, we consider first in the optimization a sum power constraint and we then show how it is possible to normalize the precoder so as to fulfill the per-TX power constraint in Sub-subsection~\ref{se:Algo:perTX}. Considering directly the per-TX power constraint requires finding the values of one Lagrangian variable per-TX, which is not practical.
\begin{remark}
In the following, we use a particular precoding algorithm to solve the optimization problem \eqref{eq:optimization_approximation}. However, our approach for the hierarchical CSI configuration can be adapted to other precoding algorithm or others figures-of-merit. The sole requirement being that having a part of the precoding coefficients fixed should not make the optimization intractable.\qed
\end{remark}

\subsubsection{Review of the Sum Rate Maximization Algorithm~\cite{Christensen2008,Negro2012,Fritzsche2013a}}
To optimize the conditional average sum rate given the precoders at TX~$k$ for $k<j$, we build upon the approach in \cite{Christensen2008} which is one of the most well known sum rate maximization algorithm. This algorithm exploits the relation between the MSE minimization and the sum rate maximization. Specifically, it is shown in \cite{Christensen2008} that a local optimum for the sum rate maximization is reached almost surely by solving the optimization problem:
\begin{equation}
\begin{aligned}
(\bG_k^{\star},\bT^{\star},\bm{\Omega}_k^{\star})&=\argmin_{\bG_k,\bT,\bm{\Omega}_k}  \sum_{k=1}^K\trace\LB \bm{\Omega}_k\bM_k\RB-\log_2|\bm{\Omega}_k| \\
&~~~~~~~~~~~~~~~~~~~~~~~~~~~\text{, s.to $\|\bT\|_{\Fro}^2\leq P$}  
\end{aligned}
\label{eq:optimization_christensen}
\end{equation}  
with $\bM_k$ being the MSE matrix defined in \eqref{eq:SM_7} and $\bm{\Omega}_k\in \mathbb{C}^{d_k\times d_k}$ being a weighting matrix left to be optimized. In fact, we will consider a robust precoding algorithm where the objective consists of the expected sum rate conditioned on the knowledge of the channel estimate at the TX. This comes down to replacing the MSE matrix~$\bM_k$ by the average MSE matrix~$\bar{\bM}_k$ defined as~\cite{Negro2012,Fritzsche2013a}
\begin{align}
\bar{\bM}_k&\triangleq \E_{\bm{\Delta}^{(j)}}[\bM_k]\\
&= \bM_k+\E[\bG_k^{\He}(\bm{\Delta}^{(j)})_k^{\He}\bT\bT^{\He}\bm{\Delta}^{(j)}_k\bG_k]\\
&=\bM_k+\sum_{\ell=1}^{N_k}\sum_{p=1}^{M_{\Tot}}\sum_{\ell'=1}^{N_k}\sum_{p'=1}^{M_{\Tot}}\E\bigg[ \bG_k^{\He} \be_{\ell} \{\bm{\Delta}^{(j)}_k\}^{*}_{p,\ell}\be_p^{\He}\bT\notag\\
&~~~~~~~~~~~~~~~~~~~~~~~~~~~~~~\bT^{\He}\be_{p'} \{\bm{\Delta}^{(j)}_k\}_{p',\ell'}\be_{\ell'}^{\He}\bG_k\bigg]\\
&=\bM_k+ \bG_k^{\He} \bm{\Phi}_k\bG_k
\label{eq:proof_prop_5}
\end{align}
with the matrix~$\bm{\Phi}_k\in \mathbb{C}^{N_k\times N_k}$ being a diagonal matrix defined such that~$\forall i \in \{1,\ldots, N_k\}$,
\begin{equation}
\begin{aligned}
\{\bm{\Phi}_k\}_{i,i}\triangleq \trace \LB \diag \LB (\sigma_{1,i}^{(j)})^2,\ldots,(\sigma_{M_{\Tot},i}^{(j)})^2\RB \diag\LB\bT\bT^{\He} \RB \RB.
\end{aligned}
\label{eq:proof_prop_6}
\end{equation} 
The objective of the minimization can also be rewritten as
\begin{equation}
\begin{aligned}
&\sum_{k=1}^K\trace \LB \bm{\Omega}_k \bar{\bM}_k \RB\\
&=\trace \LB \bm{\Omega}\bM \RB +\trace\LB \bT^{\He}\bm{\Psi} \bT\RB\\
\end{aligned}
\label{eq:proof_prop_5}
\end{equation}     
with $\bm{\Omega}\triangleq\mathrm{blockdiag}(\bm{\Omega}_1,\ldots,\bm{\Omega}_K)$ and the matrix~$\bm{\Psi}\in \mathbb{C}^{M_{\Tot}\times M_{\Tot}}$ being a diagonal matrix such that $\forall i \in \{1,\ldots, M_{\Tot}\}$,
\begin{equation}
\begin{aligned}
\{\bm{\Psi}\}_{i,i}\triangleq \trace \LB \bG  \bG^{\He} \diag \LB (\sigma^{(j)}_{i,1})^{2},\ldots,(\sigma^{(j)}_{i,N_{\Tot}})^{2} \RB\RB.
\end{aligned}
\label{eq:proof_prop_6}
\end{equation} 
This minimization is convex in each of the optimization variables. It follows that updating each of the optimization variables alternatively, the algorithm converges to a local maximum. For brevity and to focus on the specificity of the hierarchical algorithm, the details of the update of the optimization variables is relegated to Appendix~\ref{app:centralized}.

The RX filters are updated as the minimum MSE
\begin{equation}
\bG_k^{\He}=\bT_k^{\He} \bH_k \LB{\bH}_k\bT\bT^{\He}{\bH}_k^{\He}+ \bI_{N_k}+\bm{\Phi}_k\RB^{-1}.	
\label{eq:algo_5}
\end{equation}
and the weighting matrix~$\bm{\Omega}_k$ as
\begin{equation}
\bm{\Omega}_k=\bar{\bM}_k^{-1},\qquad \forall k.
\label{eq:algo_6}
\end{equation} 
Furthermore, it is shown the optimal update of~$\bT$ is given by
\begin{equation}
\bT\!=\!\LB \bH\bG\bm{\Omega}\bG^{\He}\bH^{\He}\!+\!\frac{\trace(\bm{\Omega}\bG^{\He}\bG)}{P_{\Tot}}\I_{M_{\Tot}}\!+\!\bm{\Psi}\RB^{-1}\!\!\!\!\bH \bG\bm{\Omega}.
\label{eq:algo_7}
\end{equation} 

\subsubsection{Hierarchical Sum Rate Maximization}
In the case of hierarchical precoding, it can easily be seen that the update of the RX filters and the weighting matrices is the same as in the centralized algorithm since the only difference between the two optimization problems comes from the additional constraints over the precoder~$\bT$. Indeed, the update of the precoder is different as the first $M_{\In}$ rows of~$\bT$ are fixed. For the precoder update, we take the derivative of the Lagrangian according to~$\bW_{\Opt}$ with a sum power constraint. The precoding matrix canceling this derivative is then given by \cite{Fritzsche2013b}
\begin{align}
&\bW_{\Opt}^{\He}=\LB\bH_{\Opt}\bG\bm{\Omega}\bG^{\He}\bH_{\Opt}^{\He}+\lambda \I_{M_{\Opt}}\RB^{-1}\notag\\
&~~~~~~~~~~~~~~~~ ~~~ \cdot\bH_{\Opt}\bG\bm{\Omega}\LB \I_{d_{\Tot}}- \bG ^{\He}\bH_{\In}^{\He}\bW_{\In}^{\He} \RB.
\label{eq:Algo_8}
\end{align} 
The detailed derivation of this formula is available in Appendix~\ref{app:hierarchical}. The value of~$\lambda_{\Opt}$ is then obtained by bisection to fulfill the sum power constraint. 

However, combined with the iterative update of the RX filters and the weighting matrices, this makes the algorithm relatively demanding. As an alternative, we propose to set~$\lambda_{\Opt}=0$ and simply proceed by clipping, i.e., normalizing the power used if it is larger than the power constraint. Mathematically, this is written as
\begin{equation}
\begin{aligned} 
\bW_{\Opt}^{\He}&=\frac{\bW_{\Opt}^{\He}}{\|\bW_{\Opt}\|_{\Fro}} \LB \min\LB \|\bW_{\Opt}\|_{\Fro}, \sqrt{P_{\Tot}-\|\bW_{\In}\|_{\Fro}^2}\RB\RB.
\end{aligned}
\label{eq:Algo_10}
\end{equation}

\subsubsection{Per-TX Power Constraint}\label{se:Algo:perTX}
The TXs being in fact not collocated, we consider a per-TX power constraint. It is thus necessary to scale down the precoder obtained according to the algorithm described above, to ensure that all the per-TX power constraints are fulfilled. This is obtained by setting:
\begin{equation}
\bW_{\Opt}^{\He}= \frac{\bW_{\Opt}^{\He}}{\max_{j=1}^K \|\bW_{j}\|_{\Fro}^2/P_j}.
\label{eq:Algo_9}
\end{equation}
\begin{remark}
Applying this normalization at each update of the TX has for consequence that the convergence of the algorithm is no longer guaranteed. Finding the optimal solution with the per-TX antenna power constraints taken into account is an ongoing research topic.\qed
\end{remark}


\section{Simulation Results}
To evaluate the performance of our algorithm, we average the performance over~$1000$ channel realizations via Monte-Carlo simulations. We start with a simple configuration with $K=4$~TX/RX pairs equipped each with a single antenna and with all the wireless links being unit-variance. We furthermore assume that each TX has the same power constraint. We compare our hierarchical precoding algorithm to the maximum sum rate algorithm from \cite{Christensen2008} using perfect CSIT. To fulfill the per-TX power constraint also in this perfect CSIT configuration, we normalize the precoder according to \eqref{eq:Algo_9} at each TX update. Finally, we compare our hierarchical precoding algorithm to the ``naive" \emph{distributed} use of the robust maximum sum rate algorithm from \cite{Negro2012,Fritzsche2013a} at each TX using the imperfect CSIT locally available.

\begin{figure}[htp!] 
\centering
\includegraphics[width=1\columnwidth]{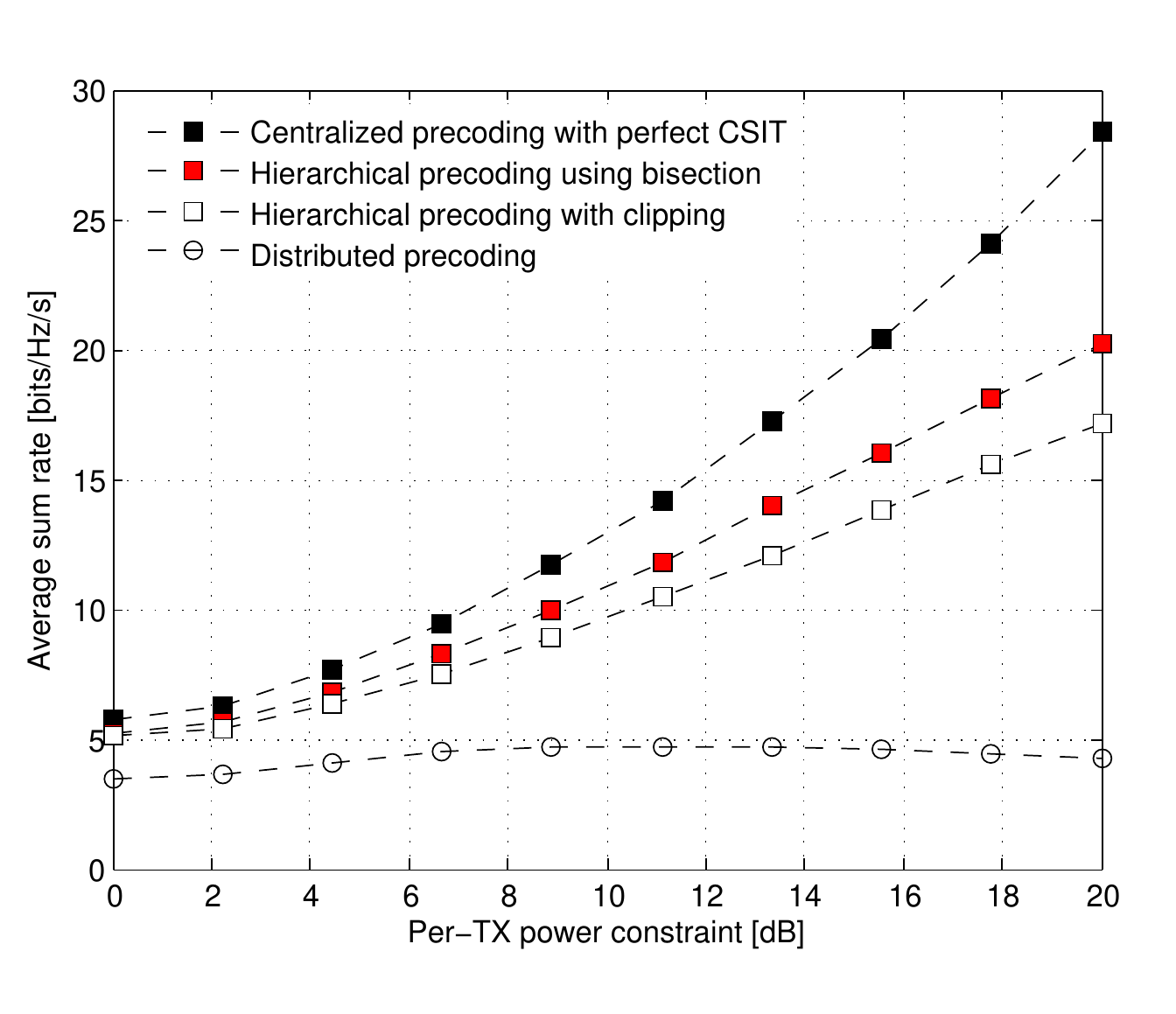}
\caption{Average sum rate as a function of the SNR~$P$.} 
\label{Rate_SPC}
\end{figure}

We show in Fig.~\ref{Rate_SPC}, the average sum rate as a function of the per-TX SNR in the following simple CSI configuration
\begin{equation}
\begin{aligned}
(\sigma_{ik}^{(j)})^2&=0.25,&\forall i,k,\qquad j=1,2\\
(\sigma_{ik}^{(j)})^2&=0,&\forall i,k,\qquad j=3,4.
\label{eq:simu_1}
\end{aligned}
\end{equation}
It can be seen that both hierarchical precoding schemes outperform very significantly the naive distributed precoding scheme. In particular, a positive DoF (slope in the SNR) is achieved. This is a consequence of TX~$3$ and TX~$4$ having perfect CSI.

\section{Conclusion}
We have developed in this work a robust precoding algorithm for the hierarchical CSI configuration which outperforms conventional precoding schemes by taking explicitly the hierarchical CSI structure into account. Finding the optimal precoder being too difficult, we have proposed an approximate solution consisting in considering during the optimization at TX~$j$ that all the TXs with a more accurate CSI have in fact received the same CSI as TX~$j$. This approach is very general and can be applied in many other wireless settings with a hierarchical information structure. 

The hierarchical CSI structure is particularly relevant for the next generation wireless networks with a partially centralized architecture and an heterogeneous backhaul. How to approach the optimal precoder in this hierarchical CSI configuration is a challenging topic which will be tackled in the future. Another interesting direction of research consists in evaluating analytically the performance obtained with a hierarchical architecture.

\FloatBarrier
\appendices

\section{Robust Precoding Algorithm in the Centralized Case \cite{Christensen2008,Negro2012,Fritzsche2013a}}\label{app:centralized}
We consider the optimization problem:
\begin{equation}
\begin{aligned}
(\bG_k^{\star},\bT^{\star},\bm{\Omega}_k^{\star})&=\argmin  \sum_{k=1}^K\trace\LB \bm{\Omega}_k\bM_k\RB-\log|\bm{\Omega}_k| \\
&~~~~~~~~~~~~~~~~~~~~~~~~~~~\text{, s.to $\|\bT\|_{\Fro}^2\leq P$}  
\end{aligned}
\label{eq:optimization_appendix}
\end{equation}  
where the objective can be written explicitly from \eqref{eq:proof_prop_5} and \eqref{eq:SM_7} as
\begin{align}
& \sum_{k=1}^K \trace \LB\bm{\Omega}_k\bar{\bM}_k\RB-\log_2|\bm{\Omega}_k| \notag\\
&\!=\!\sum_{k=1}^K \trace \LB\bm{\Omega}_k\!+\!\bm{\Omega}_k\bG_k^{\He}\bG_k\!+\!\bm{\Omega}_k\bG_k^{\He}\bH^{\He}_k\bT\bT^{\He}\bH_k\bG_k\!-\!\bm{\Omega}_k\bG_k^{\He}\bH_k^{\He}\bT_{k}\!-\!\bm{\Omega}_k\bT_k^{\He}\bH_k\bG_{k}\!+ \bm{\Omega}_k\!\bG_k^{\He} \bm{\Phi}_k\bG_k\RB\notag \\
&~~~~~~~~~~~~~~~~~~~~~~~~~~~~~~~~~~~~~~~~~~~~~~~~~~~~~~~~~~~~~~~~~~~~~~~~~~~~~~~~~~~~ -\log_2|\bm{\Omega}_k|
\label{eq:proof_Lagrangian_1} 
\end{align}
with the diagonal matrix~$\bm{\Phi}_k\in \mathbb{C}^{N_k\times N_k}$ being defined such that 
\begin{equation}
\begin{aligned}
\{\bm{\Phi}_k\}_{i,i}\triangleq \trace \LB \diag \LB (\sigma_{1,i}^{(j)})^2,\ldots,(\sigma_{M_{\Tot},i}^{(j)})^2\RB \diag\LB\bT\bT^{\He} \RB \RB,\qquad \qquad \forall i \in \{1,\ldots, N_k\}.
\end{aligned}
\label{eq:proof_Lagrangian_2} 
\end{equation} 
Taking the derivative according to~$\bG_k$ and setting it to zero, if follows easily that
\begin{equation}
\bG_k^{\He}=\bT_k^{\He} \bH_k \LB{\bH}_k\bT\bT^{\He}{\bH}_k^{\He}+ \bI_{N_k}+\bm{\Phi}_k\RB^{-1}.	
\label{eq:proof_Lagrangian_3} 
\end{equation}
In addition, taking the derivative of the objective according to~$\bm{\Omega}_k$ and setting it to zero we obtain
\begin{equation}
\bm{\Omega}_k=\bar{\bM}_k^{-1} 
\label{eq:proof_Lagrangian_4} 
\end{equation}
where we have used the following derivative formula\cite{Petersen2012}
\begin{equation}
\partial \log \LB \det \LB \bX \RB\RB =\trace \LB \bX^{-1}\partial \bX \RB.
\label{eq:proof_Lagrangian_5} 
\end{equation}
Finally, to update the precoder~$\bT$ with the sum power constraint, we define the Lagrangian, denoted by~$\mathcal{L}(\bT,\lambda)$, as
\begin{align}
\mathcal{L}(\bT,\lambda)&\triangleq \sum_{k=1}^K \trace \LB\bm{\Omega}_k\bar{\bM}_k\RB+\lambda \LB \trace\LB \bT\bT^{\He}\RB-P_{\Tot}\RB \\
&=\sum_{k=1}^K \trace \LB\bm{\Omega}_k+\bm{\Omega}_k\bG_k^{\He}\bG_k+\bm{\Omega}_k\bG_k^{\He}\bH^{\He}_k\bT\bT^{\He}\bH_k\bG_k-\bm{\Omega}_k\bG_k^{\He}\bH_k^{\He}\bT_{k}\notag\right.\\
&~~~~~~~~~~~~~~~~~~~~~~~ \left.-\bm{\Omega}_k\bT_k^{\He}\bH_k\bG_{k}+ \bG_k^{\He} \bm{\Phi}_k\bG_k\RB+\lambda \LB \trace\LB \bT\bT^{\He}\RB-P_{\Tot}\RB
\label{eq:proof_Lagrangian_6} 
\end{align}
Taking the derivative of the Lagrangian yields 
\begin{align}
\frac{\partial \mathcal{L}(\bT,\lambda)}{\partial \bT^{*}_{ji}}&=\sum_{k=1}^K\trace\LB \bm{\Omega}_k\bG_k^{\He}\bH^{\He}_k\bT\be_i\be_j^{\He}\bH_k\bG_k\RB-\trace\LB \bm{\Omega}_k\be_i\be_j^{\He}\bH_k\bG_{k}\RB\notag\\
&~~~~~~~~~~~~~~ +\sum_{\ell=1}^{N_k}\sum_{p=1}^{M_{\Tot}} \trace\LB\bm{\Omega}_k\bG_k^{\He} \be_{\ell} (\sigma'^{(j)}_{p,\ell})^{2}\be_p^{\He}\bT\be_{i}\be_j^{\He}\be_{p} \be_{\ell}^{\He}\bG_k \RB +\lambda \trace\LB \bT\be_i\be_j^{\He}\RB\\
&\!=\!\sum_{k=1}^K\be_j^{\He}\bH_k\bG_k\bm{\Omega}_k\bG_k^{\He}\bH^{\He}_k\bT\be_i\!-\!\be_j^{\He}\bH_k\bG_{k}\bm{\Omega}_k\be_i\!+\! \sum_{\ell=1}^{N_k} \be_{\ell}^{\He}\bG_k\bm{\Omega}_k\bG_k^{\He}  (\sigma'^{(j)}_{j,\ell})^{2}\be_{\ell}\be_j^{\He}\bT\be_{i} \notag\\
&~~~~~~~~~~~~~~~~~~~~~~~~~~~~~~~~~~~~~~~~~~~~~~~~~~~~~~~~~~~~~~~~~~~~~~~~~~~+\!\lambda \be_j^{\He}\bT\be_i.
\label{eq:proof_Lagrangian_7}
\end{align} 
Taking the derivative according to all the elements of~$\bT^{*}$, we obtain
\begin{align}
\frac{\partial \mathcal{L}(\bT,\lambda)}{\partial \bT^{*}}
&=\bH\bG\bm{\Omega}\bG^{\He}\bH^{\He}\bT-\bH\bG\bm{\Omega}+\bm{\Psi} \bT+\lambda \bT 
\label{eq:proof_Lagrangian_8}
\end{align} 
with the diagonal matrix~$\bm{\Psi}\in \mathbb{C}^{M_{\Tot}\times M_{\Tot}}$ being defined such that 
\begin{equation}
\begin{aligned}
\{\bm{\Psi}\}_{i,i}\triangleq \trace \LB \bG  \bG^{\He} \diag \LB (\sigma^{(j)}_{i,1})^{2},\ldots,(\sigma^{(j)}_{i,N_{\Tot}})^{2} \RB\RB,\qquad \qquad \forall i \in \{1,\ldots, M_{\Tot}\}.
\end{aligned}
\label{eq:proof_Lagrangian_9}
\end{equation}  
The precoder which cancels the derivative of the Lagrangian is then given by
\begin{equation}
\begin{aligned}
\bT&=\LB \bH\bG\bm{\Omega}\bG^{\He}\bH^{\He}  +\bm{\Psi}+\lambda \I_{M_{\Tot}} \RB^{-1} \bH \bG\bm{\Omega}.
\end{aligned}
\label{eq:proof_Lagrangian_10}
\end{equation}
It remains then solely to calculate the value of the Lagrangian variable~$\lambda$. Following the approach of \cite{Joham2002}, we scale the RX filter $\bG^{\He}$ by the positive scalar~$\beta^{-1}$. Using \eqref{eq:SM_7} and \eqref{eq:proof_prop_5}, the weighted MSE is then written as 
\begin{align}
\sum_{k=1}^K \trace \LB\bm{\Omega}_k\bar{\bM}_k \RB
&=\sum_{k=1}^K \trace \LB\bm{\Omega}_k+\beta^{-2}\bm{\Omega}_k\bG_k^{\He}\bG_k+\beta^{-2}\bm{\Omega}_k\bG_k^{\He}\bm{\Phi}_k\bG_k+\beta^{-2}\bm{\Omega}_k\bG_k^{\He}\bH^{\He}_k\bT\bT^{\He}\bH_k\bG_k\right.\notag\\
&\left.~~~~~~~~~~~~~~~~~~~~~~~~~~~~~~~~~~~~~~-\beta^{-1}\bm{\Omega}_k\bG_k^{\He}\bH_k^{\He}\bT_{k}-\beta^{-1}\bm{\Omega}_k\bT_k^{\He}\bH_k\bG_{k} \RB\\
&= \trace \LB\bm{\Omega} +\beta^{-2}\bm{\Omega}\bG^{\He}\bG+\beta^{-2}\bm{\Omega}\bG^{\He}\bm{\Phi}\bG+\beta^{-2}\bm{\Omega} \bG^{\He}\bH^{\He}\bT\bT^{\He}\bH\bG\right.\notag\\
&\left.~~~~~~~~~~~~~~~~~~~~~~~~~~~~~~~~~~~~~~~~~~~~~~~~~~~~~~~~~-2\beta^{-1}\Re\LB \bm{\Omega}\bG^{\He}\bH^{\He}\bT\RB\RB
\label{eq:proof_Lagrangian_11}
\end{align}  
Taking the derivative of the Lagrangian according to~$\beta$ gives
\begin{equation}
\frac{\partial \mathcal{L}(\bT,\lambda)}{\partial \beta}\!=\!-2\beta^{-2}\trace \LB \beta^{-1}\bm{\Omega}\bG^{\He}\bG\!+\!\beta^{-1}\bm{\Omega}\bG^{\He}\bm{\Phi}\bG\!+\!\beta^{-1}\bm{\Omega}\bG^{\He}\bH^{\He}\bT\bT^{\He}\bH\bG-\Re\LB \bm{\Omega}\bG^{\He}\bH^{\He}\bT\RB \RB
\label{eq:proof_Lagrangian_12}
\end{equation} 
Inserting then the precoder before normalization~$\tilde{\bT}\triangleq \beta^{-1}\bT$ such that $\bT=\beta\tilde{\bT}$ gives
\begin{equation}
\frac{\partial \mathcal{L}(\bT,\lambda)}{\partial \beta}\!=\!-2\beta^{-3}\trace \LB \bm{\Omega}\bG^{\He}\bG\!+\! \bm{\Omega}\bG^{\He}\bm{\Phi}\bG\!+\!\beta^{2}\bm{\Omega}\bG^{\He}\bH^{\He}\tilde{\bT}\tilde{\bT}^{\He}\bH\bG\!-\!\beta^{2}\Re\LB \bm{\Omega}\bG^{\He}\bH^{\He}\tilde{\bT}\RB \RB
\label{eq:proof_Lagrangian_13}
\end{equation} 
We can then rewrite the last term as  
\begin{align}
&\Re\LB\trace\LB \bm{\Omega}\bG^{\He}\bH^{\He}\bT\RB\RB\notag\\
&=\Re\LB\trace\LB \bm{\Omega}\bG^{\He}\bH^{\He}\LB \bH\bG\bm{\Omega}\bG^{\He}\bH^{\He}+\bm{\Psi}+\lambda\I_{M_{\Tot}}\RB^{-1}\LB \bH\bG\bm{\Omega}\bG^{\He}\bH^{\He}+\bm{\Psi}+\lambda\I_{M_{\Tot}}\RB \tilde{\bT}\RB\RB\\ 
&\stackrel{(a)}{=}\Re\LB \trace\LB \tilde{\bT}^{\He}\LB \bH\bG\bm{\Omega}\bG^{\He}\bH^{\He}+\bm{\Psi}+\lambda\I_{M_{\Tot}}\RB \tilde{\bT}\RB\RB\\ 
&\stackrel{(b)}{=} \trace\LB\tilde{\bT}^{\He}\LB \bH\bG\bm{\Omega}\bG^{\He}\bH^{\He}+\bm{\Psi}+\lambda\I_{M_{\Tot}}\RB \tilde{\bT}\RB 
\label{eq:proof_Lagrangian_14}
\end{align} 
where equality~$(a)$ results from the expression of~$\bT$ in \eqref{eq:proof_Lagrangian_4} and the fact that the matrix~$ \bm{\Omega}$ can be shown to be hermitian. Equality~$(b)$ holds because the matrix inside the trace is Hermitian. Inserting \eqref{eq:proof_Lagrangian_14} inside \eqref{eq:proof_Lagrangian_13} gives
\begin{align}
\frac{\partial \mathcal{L}(\bT,\lambda)}{\partial \beta}&=-2\beta^{-3}\trace \LB \bm{\Omega}\bG^{\He}\bG+\bm{\Omega}\bG^{\He}\bG-\beta^{2}   \tilde{\bT}^{\He}\LB\bm{\Psi}+\lambda\I_{M_{\Tot}}\RB \tilde{\bT}  \RB\\
&=-2\beta^{-3}\trace \LB \bm{\Omega}\bG^{\He}\bG+ \bm{\Omega}\bG^{\He}\bm{\Phi}\bG- \bT^{\He} \bm{\Psi}\bT+\lambda\bT^{\He}\bT   \RB\\
&=-2\beta^{-3}\LB \trace \LB \bm{\Omega}\bG^{\He}\bG \RB+ \lambda \trace \LB\bT^{\He}\bT \RB\RB
\label{eq:proof_Lagrangian_15}
\end{align}
where the last equality is satisfied because we have by definition that
\begin{equation}
\trace \LB\bm{\Omega}\bG^{\He}\bm{\Phi}\bG\RB=\trace \LB \bT^{\He}\bm{\Psi}\bT\RB.
\label{eq:proof_Lagrangian_16}
\end{equation}
Setting the derivative in \eqref{eq:proof_Lagrangian_15} to zero, and solving for $\lambda$ leads to
\begin{align}
\lambda&=\frac{\trace \LB \bm{\Omega}\bG^{\He}\bG \RB}{\trace \LB\bT^{\He}\bT\RB}\\
&=\frac{\trace \LB \bm{\Omega}\bG^{\He}\bG \RB}{P_{\Tot}}
\label{eq:proof_Lagrangian_17}
\end{align} 
which gives the final expression

\section{Robust Precoding Algorithm with Hierarchical CSIT}\label{app:hierarchical}
We consider then the optimization problem:
\begin{equation}
\begin{aligned}
(\bG_k^{\star},\bT^{\star},\bm{\Omega}_k^{\star})&=\argmin  \sum_{k=1}^K\trace\LB \bm{\Omega}_k\bar{\bM}_k\RB-\log_2|\bm{\Omega}_k| \\
&~~~~~~~~~~~~~~~~~~~~~~~~~~~\text{, s. to~$\|\bT\|_{\Fro}^2\leq P$}\\  
&~~~~~~~~~~~~~~~~~~~~~~~~~~~\text{, s. to~$\bW_{i}^{\He}=(\bX^0_{i})^{\He},\qquad \forall i=1,\ldots,j-1$.}  
\end{aligned}
\label{eq:optimization_christensen}
\end{equation}  
The only difference with the optimization in the centralized case comes from the additional constraint over the precoder~$\bT$. Hence, the update of the RX filter~$\bG_k$ and the update of the weighting matrix~$\bm{\Omega}_k$ remain the same, and only the update of the precoder has to be modified. In fact, the derivative of the Lagrangian is obtained by applying the expression obtained in \eqref{eq:proof_Lagrangian_7} for the elements of~$\bW_{\Opt}^{\He}$. This then gives 
\begin{equation}
\begin{aligned}
\bE_{\Opt}^{\He} \bH \bG\bm{\Omega}\bG^{\He}\bH ^{\He} \bT -\bE_{\Opt}^{\He}\bH \bG\bm{\Omega}  +\bE_{\Opt}^{\He} \bm{\Psi}\bT+\lambda\bE_{\Opt}^{\He} \bT &= \bm{0}.
\end{aligned}
\end{equation}
Writing~$\bT$ as a function of~$\bW_{\In}^{\He}$ and~$\bW_{\Opt}^{\He}$ in the first term and using that the matrix~$\bm{\Psi}$ is diagonal, we obtain the update of the precoder given in \eqref{eq:Algo_10}.

\bibliographystyle{IEEEtran}
\bibliography{./../../Literatur}
\end{document}